\documentclass{PoS}
\newcommand{\RA}{RadioAstron}
\newcommand{\tb}{{$\mathrm{T_B }$}}
\newcommand{\uas}{$\mu$as}

\title{Interstellar Scintillation Monitoring of the \RA\ Blazars}
\ShortTitle{ISS Monitoring of \RA\ Blazars}

\author{
{Jun Liu $^{1,2,3,}$\thanks{speaker}, 
Thomas P. Krichbaum$^{1}$, 
Hayley Bignall$^{4}$, 
Xiang Liu $^{2,3}$, 
Alex Kraus$^{1}$, 
Yuri Y. Kovalev$^{5,6,1}$, 
Kirill V. Sokolovsky$^{7,8,5}$,
Giuseppe Cim\`{o}$^{9,10}$,
J. Anton Zensus$^{1}$}\\
\llap{$^1$} Max-Planck-Institut f\"ur Radioastronomie, Auf dem H\"ugel 69, D-53121 Bonn, Germany\\
\llap{$^2$} Xinjiang Astronomical Observatory, CAS, 150 Science 1-Street, Urumqi 830011, P. R. China\\
\llap{$^3$} Key Laboratory of Radio Astronomy, CAS, Urumqi 830011, P. R. China\\
\llap{$^4$} CSIRO Astronomy and Space Science, PO Box 1130, Bentley WA 6102, Australia \\
\llap{$^5$} Astro Space Center of Lebedev Physical Institute, Profsoyuznaya 84/32, 117997 Moscow, Russia \\
\llap{$^6$} Moscow Institute of Physics and Technology, Dolgoprudny, Institutsky per., 9, Moscow region, 141700, Russia \\
\llap{$^7$} Physics and Astronomy Department, Michigan State University, East Lansing, MI 48824, USA \\
\llap{$^8$} Sternberg Astronomical Ins., Moscow University, Universitetskii pr. 13, 119992 Moscow, Russia \\
\llap{$^9$} Joint Institute for VLBI ERIC, P.O. Box 2, 7990 AA Dwingeloo, The Netherlands\\
\llap{$^{10}$} ASTRON, Netherlands Institute for Radio Astronomy, P.O. Box 2, 7990 AA Dwingeloo, The Netherlands\\
E-mail: \email{jliu@mpifr-bonn.mpg.de}}

\abstract{The \RA\ space radio telescope provides a unique opportunity to study the extreme brightness temperatures (\tb) in AGNs with unprecedented long baselines of up to 28 Earth diameters. Since interstellar scintillation (ISS) may affect the visibilities observed with space VLBI (sVLBI), a complementary ground based flux density monitoring of the \RA\ targets, which is performed near in time to the VLBI observation, could be beneficial. The combination/comparison with the sVLBI data can help to unravel the relative influence of source intrinsic and ISS induced effects, which in the end may alter the conclusions on the \tb\ measurements from sVLBI. Since 2013, a dedicated monitoring program has been ongoing to observe the ISS of \RA\ AGN targets with a number of radio telescopes. Here we briefly introduce the program and present results from the statistical analysis of the Effelsberg monitoring data. We discuss the possible effects of ISS on \tb\ measurements for the \RA\ target B0529+483 as a case study.}

\FullConference{14th European VLBI Network Symposium \& Users Meeting (EVN 2018)\\
		8-11 October 2018\\
		Granada, Spain}

\begin{document}

\vspace{-0.3cm}
\section{Introduction}
\label{sec:intro}
\vspace{-0.3cm}
\RA\ is an international collaborative mission with a 10-m radio telescope onboard the SPEKTR-R spacecraft launched on July 18, 2011 (\cite{Kovalev2012}). With unprecedented long baselines of up to 28 Earth diameters (EDs), it is capable of measuring apparent brightness temperatures (at frequencies 0.3--22\,GHz) directly up to extreme \tb\ values of $10^{15}$-$10^{16}$\,K. 
The \RA\ AGN survey has revealed dozens of sources with detectable fluxes at $>10$ EDs and \tb\ well excess of the inverse-Compton limit (\cite{Kovalev2018}). The extreme high \tb\ sources detected by \RA\ are so compact that they should scintillate. As a consequence the space VLBI (sVLBI) visibilities and \tb\ measurements are believed to be affected by scintillation, especially on long baselines.

In order to account for the contribution of ISS during the \RA\ observations, a dedicated program to monitor the \RA\ AGN survey targets has been conducted with a number of radio telescopes on Earth: ATCA, Effelsberg, WSRT, Urumqi and OVRO. When logistically feasible, the ground based ISS monitoring is performed close in time to the sVLBI observations. This offers an independent probe of the structure of blazar cores on \uas\ angular scales. The direct measurement of the sizes of scintillating sources with \RA\ can help to determine the properties of the ISS screens, such as their distance and scattering strength. In turn, the focusing and defocusing effects caused by ISS and being monitored from ground may have a significant influence on the measured sVLBI visibilities. For a comprehensive analysis of the \RA\ data,
it is essential to better understand these effects. Thus, these two independent probes of \uas-scale structure are highly complementary.

In the following, we present the statistical results of the variability analysis from the Effelsberg monitoring at 4.85\,GHz. We then compare the \tb\ derived from ISS modeling and sVLBI observations and discuss a possible scattering scenario for the quasar B0529+483.

\vspace{-0.3cm}
\section{Observations, Data Calibration and Statistical Results}
\label{sec:obs}
\vspace{-0.3cm}
So far, seven sessions of monitoring have been performed with the Effelsberg 100m radio telescope at 4.85\,GHz. For each observing session, the main targets were chosen from the \RA\ block schedule and observed along with a number of primary and secondary calibrators. All the observations were performed in cross scan mode, with typical integration time of $\sim$ 1.5 minutes per source. The data calibration follows the well established, standard procedure 
(\cite{Liu2018, Kraus2003}). The basic observational details are summarized in Table~\ref{tab:obs_info}. 
An example of calibrated lightcurves for B0529+483 is shown in the left panel of Figure~\ref{fig:lightcurve}. 

\begin{table}[ht]
\vspace{-0.2cm}
\setlength{\belowcaptionskip}{0.2cm}
\centering
\caption{Basic information for the seven epochs of observing sessions. "Sampling" is the mean number of flux density measurements per hour, "Duty Cycle" the average number of measurements per hour for each source, and $m_c$ the average fractional variability of flux density of the calibrators in percent.}
\label{tab:obs_info}
\scalebox{0.9}[0.85]{
\begin{tabular}{|p{1cm}|p{3.5cm}|p{1.5cm}|p{1cm}|p{2cm}|p{2cm}|p{1cm}|}
\hline
Epoch & Date  & Duration    & Source  & Sampling  & Duty Cycle     & $m_c$ \\
 &   &  [{$h$}] & Num.  &  [$h^{-1}$]  & [$h^{-1}$]  & [{\%}]  \\
\hline
A14 & 18.07--20.07.2014     & 62.0   & 37    & 14.8    & 0.40    & 0.50  \\
B14 & 12.09--15.09.2014     & 66.6   & 45    & 15.9    & 0.35    & 0.40  \\
A15 & 31.07--06.08.2015     & 73.6   & 42    & 14.3    & 0.34    & 0.40  \\
B15 & 17.12--21.12.2015     & 82.4   & 39    & 14.5    & 0.37    & 0.35  \\
A16 & 20.12--24.12.2016     & 84.4   & 41    & 14.0    & 0.34    & 0.60  \\
A17 & 04.10--08.10.2017     & 83.9   & 39    & 14.6    & 0.35    & 0.60  \\
A18 & 31.03--03.04.2018     & 89.1   & 39    & 14.1    & 0.34    & 0.50  \\
\hline
\end{tabular}}
\end{table}

Of the 161 targets observed, 52 sources exhibited rapid intra-day variability (IDV) in at least one observing epoch. This leads to an IDV detection rate of $\sim$30\% in our monitoring sample, consistent with earlier studies (e.g. \cite{Lovell2008}). Of the 23 \RA\ AGN survey targets in our sample, 12 showed IDV, indicating that half of the \RA\ targets are scintillating, and caution should be taken when interpreting their brightness temperatures derived from sVLBI.

\vspace{-0.3cm}
\section{Independent Probe of \tb\ with ISS Monitoring-- the Case of B0529+483}
\label{sec:b0529}
\vspace{-0.3cm}

Pilipenko et al. (2018,~\cite{Pilipenko2018}) reported the \RA\ observations of blazar B0529+483 between Sep. 2012 and Nov. 2014. At both 4.8\,GHz and 22\,GHz, \tb\ $\sim 10^{13}$\,K is indicated. However the observations are still inadequate to unambiguously disentangle between source intrinsic and extrinsic scattering contributions, due to simplicity of the NE2001 model (\cite{Cordes2002}) and sparse $uv$-coverage of the sVLBI data on precise angular size determination. As a result, the authors can neither confirm nor reject the impact of refractive scattering on space-ground baseline detections.

The scattering properties can be alternatively investigated using either a power spectrum (e.g. \cite{Macquart2006}) or structure function (e.g. \cite{Liu2015}) modeling of the densely sampled ISS lightcurves. At Effelsberg, significant IDV of the blazar B0529+483 was detected in three sessions out of four sessions (A14: no, A16: yes, A17: yes, A18: yes; see Table~\ref{tab:obs_info}) in which it was observed. The source shows significant long-term flux variability. Following the flare in Oct. 2012 - May 2013 \cite{Pilipenko2018}, its decay may coincide with the disappearance of the compact component, leading to no IDV detected in the low flux state in Jul. 2014, and possibly explaining the \RA\ non-detections in Sep. 2013 and Nov. 2014. A subsequent rise seen in our data sees the return of IDV in late 2016, indicating re-brightening of the compact structure.

\begin{figure*}
\vspace{-0.2cm}
\setlength{\abovecaptionskip}{-0.1cm}
    \centering
    \includegraphics[height=4.5cm]{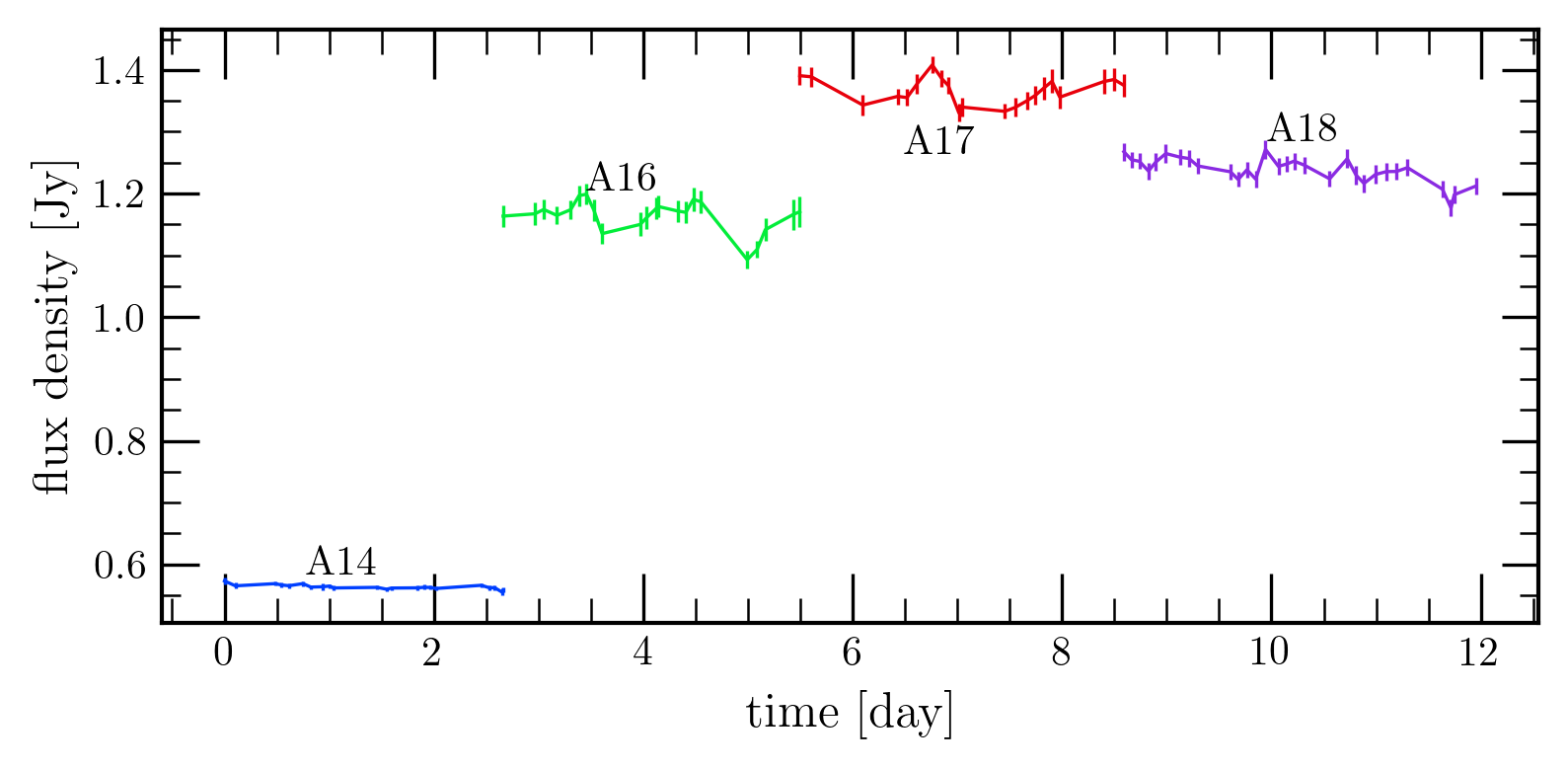}
    \includegraphics[height=4.5cm]{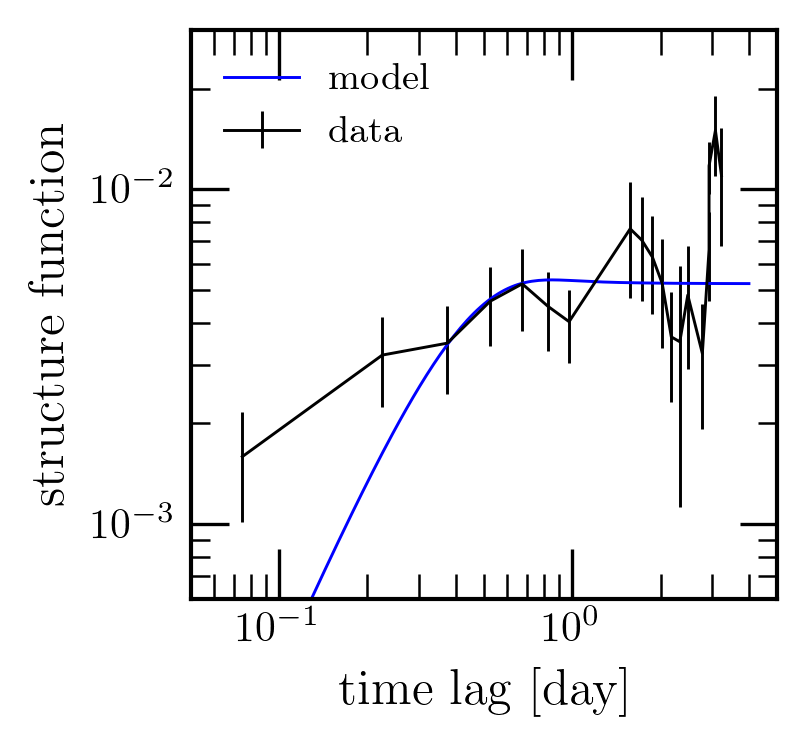}
    \caption{Left: Multiple epoch lightcurves for B0529+483 at 4.85\,GHz, for each epoch the x-axis is shifted for better visualization. Right: Model fitting of the stacked structure function.}
    \label{fig:lightcurve}
\vspace{-0.3cm}
\end{figure*}

The multi-epoch structure functions are calculated with a time binning of $\sim$ 0.15 days and then stacked in order to achieve a better signal-to-noise ratio. Following \cite{Liu2015} an analytic model is fitted to the stacked structure function with a fixed value of V$_\bot$\,=\,50km/s (characteristic relative transverse velocity between observer and scattering screen). The best fit (see right panel of Figure~\ref{fig:lightcurve}) gives a scattering measure of SM\,=\,$5.3\pm6.7\times10^{-3}\,\mathrm{kpc}\;\mathrm{m}^{-20/3}$ and a distance to the scattering screen of D\,=\,$0.31\pm0.18$\,kpc. 

With Equation (7) from \cite{Liu2015}, the combination of V$_\bot$ and D yields an apparent angular source size of $\theta_a$\,=\,27\,\uas. This value is in good agreement with the angular size of the second Gaussian component (18$\pm$15\uas) at 4.8\,GHz, supporting that there are no refractive scattering effects in B0529+483 observations with \RA. Although the possibility of `weak angular broadening' still remains, the slightly smaller value derived from ISS monitoring (comparing to the $\mathrm{\theta_{scatt}=40}$\,\uas\ in \cite{Pilipenko2018}) indicates even `weaker' angular broadening in this source. Thus our result seems to suggest that \RA\ revealed the true source structure on baselines of $\sim$20 EDs for B0529+483 . Furthermore, by assuming the core containing the variable flux only/total flux, we're able to calculate the lower/upper limit of \tb, respectively. With S\,=\,$1.252\pm0.022$\,Jy during the epochs when the source showed IDV, its \tb\ must fall in the range of [0.35, 19.6]\,$\mathrm{\times10^{13}}$\,K, which is still realistic for an assumed AGN-typical moderate Doppler boosting ($\delta\leq 20$).

\vspace{-0.3cm}
\section{Summary}
\vspace{-0.3cm}
We presented new results from a ground based ISS monitoring program in support for the \RA\ AGN survey KSP. We summarized statistical results from the Effelsberg monitoring at 4.85\,GHz. We showed that 52\% of the \RA\ AGN targets exhibited significant IDV in at least one epoch, indicating that scattering effect plays a role in these sources. 

We analyzed the scintillation of blazar B0529+483, which showed IDV during Dec. 2016, Oct. 2017 and Apr. 2018. Our result suggests an intrinsic angular size of $\sim$\,27\,\uas\ and an apparent brightness temperature \tb\ in the range [0.35, 19.6]\,$\mathrm{\times10^{13}}$\,K in this source. The consistency between single dish ISS monitoring and sVLBI observations supports the view that the former is a powerful proxy for AGN cores on scales of a few ten \uas.

\vspace{-0.3cm}
\acknowledgments{\vspace{-0.3cm}
This paper made use of data obtained with the 100-m telescope of the MPIfR at Effelsberg. This research was partially supported by the National Key R\&D Program of China (No. 2018YFA0404602), the 973 Program (No. 2015CB857100), the National Natural Science Foundation of China (No. U1731103, 11503071). YYK acknowledges support by the government of the Russian Federation (No. 05.Y09.21.0018) and the Alexander von Humboldt Foundation.}

\end{document}